\documentclass[conference]{IEEEtran}

\ifCLASSINFOpdf
\else
\fi
\usepackage{url}

\usepackage[]{fancyhdr} %
\newcommand{\changefont}{\fontsize{9}{9}\selectfont}
\usepackage[utf8]{inputenc}
\usepackage[english]{babel}
\usepackage{comment,url}

\usepackage{amssymb,amsmath}
\usepackage{mathtools}
\usepackage{graphicx}
\usepackage{verbatim}
\usepackage{wrapfig}
\usepackage{caption}
\usepackage{subcaption}
\usepackage[T1]{fontenc}
\DeclareMathAlphabet{\altmathcal}{OMS}{cmsy}{m}{n}
\renewcommand{\mathcal}{\altmathcal}
\usepackage{bm,bbm}
\usepackage{multicol}

\newcommand{\ba}[1]{\begin{array}{#1}}
\newcommand{\ea}{\end{array}}

\newcommand{\be}{\begin{equation}}
\newcommand{\ee}{\end{equation}}
\newcommand{\bea}{\begin{eqnarray*}}
\newcommand{\eea}{\end{eqnarray*}}

\newcommand{\C}{\mathcal{C}}

\newcommand{\ptdf}{\Phi}
\newcommand{\lmp}{{\rm LMP}}

\renewcommand{\doteq}{\vcentcolon=}

\usepackage[dvipsnames]{xcolor}
\usepackage{xspace}

\IEEEoverridecommandlockouts
\begin{document}

%
\title{Machine Learning for Electricity Market Clearing}


\author{
\IEEEauthorblockN{Laurent Pagnier \IEEEauthorrefmark{1}, Robert Ferrando \IEEEauthorrefmark{1}, Yury Dvorkin  \IEEEauthorrefmark{2},
and Michael Chertkov \IEEEauthorrefmark{1}
\IEEEauthorblockA{\IEEEauthorrefmark{1} Program in Applied Mathematics, 
			University of Arizona, 
			Tucson, AZ, USA\\
			\{laurentpagnier,rferrando,chertkov\}@math.arizona.edu}
\IEEEauthorblockA{\IEEEauthorrefmark{2} Electrical and Computer Engineering, New York University,
		New York, NY, USA\\
			dvorkin@nyu.edu}
}
}


%



\lhead{ACCEPTED FOR PRESENTATION IN 11TH BULK POWER SYSTEMS DYNAMICS AND CONTROL SYMPOSIUM (IREP 2022), JULY 25-30, 2022, BANFF, CANADA}


\maketitle
\thispagestyle{fancy}
\pagestyle{fancy}


\begin{abstract}
This paper seeks to design a machine learning twin of the optimal power flow (OPF) optimization, which is used in market-clearing procedures by wholesale electricity markets. The motivation for the proposed approach stems from the need to obtain the digital twin, which is much faster than the original, while also being sufficiently accurate and producing consistent generation dispatches and locational marginal prices (LMPs), which are primal and dual solutions of the OPF optimization, respectively.  Availability of market-clearing tools based on this approach will enable computationally tractable evaluation of multiple dispatch scenarios under a given unit commitment. Rather than direct solution of OPF, the Karush–Kuhn–Tucker (KKT) conditions for the OPF problem in question may be written, and in parallel the LMPs of generators and loads may be expressed in terms of the OPF Lagrangian multipliers. Also, taking advantage of the practical fact that many of the Lagrangian multipliers associated with lines will be zero (thermal limits are not binding), we build and train an ML scheme which maps flexible resources (loads and renewables) to the binding lines, and supplement it with an efficient power-grid aware linear map to optimal dispatch and LMPs. The scheme is validated and illustrated on IEEE models. We also report a trade of analysis between quality of the reconstruction and number of samples needed to train the model. 
\end{abstract}



%
\IEEEpeerreviewmaketitle

\section{Introduction}

This paper applies machine learning for accelerating computations of market-clearing outcomes in short-term electricity markets, such as hour-ahead or real-time dispatches and locational marginal prices (LMPs), with the intention of enabling high-speed and high-accuracy computations for a large number of net load (load minus renewables) conditions. In particular, we are motivated by the physics-informed machine learning (PIML) schemes outlined in \cite{Laurent2}-\cite{Laurent1} for parameter and state estimation of large-scale power transmission networks, and seek to extend these approaches to enable accurate and \textit{coherent} prediction of power dispatches and LMPs. The need for coherence arises from the requirement for dispatch and LMP predictions to satisfy market design properties (e.g. efficiency, cost recovery, and revenue adequacy, \cite{Varian}), which is a particular challenge for emerging stochastic market designs \cite{YuryCC}.  To this end, we seek to relate primal (dispatch) and dual (LMPs) Optimal Power Flow (OPF) solutions by internalizing conditions for market efficiency, cost recovery, and revenue adequacy in the proposed machine learning approach.  Therefore, we develop a variant of the ``classification-then-regression'' (CTR) method proposed in \cite{Pascal2021}. CTR is used to predict generator dispatch, governed by SCED, through regression, after employing classification to discern which generators are ``saturated,'' i.e., functioning at their minimum or maximum capacities. We intend to follow the CTR framework for power lines, determining which lines are saturated at their thermal limits. These thermal limits, in addition to other constraints, can be encapsulated via the Karush-Kuhn-Tucker (KKT) conditions (see, e.g., \cite{beck2014introduction}), which are necessary for optimality of the solution (and sufficient under conditions satisfied under DC-OPF assumptions) and relate both primal and dual variables. By the KKT condition of complementary slackness, lines which are not saturated should have a corresponding dual variable of zero. In addition, as will be derived below, and is discussed in \cite{Litvinov}, LMPs can be computed directly via a system of equations relying on the KKT conditions. As such, CTR will enable us to quickly solve for the variables needed to predict LMPs, representing an improvement in how LMPs are computed via ML by relating them to primal (dispatch) solutions using the market efficiency, cost recovery, and revenue adequacy requirements. The method outlined will allow system operators and market participants to quickly ascertain risk, and respond with strategic, cost and profit-effective operational decisions and bids. 

In addition to the  PIML schemes for power systems in \cite{Laurent2}-\cite{Laurent1}, ML techniques have become increasingly prominent in enhancing the solution of optimal power flow problems. The impetus for injecting ML into OPF is almost universally to improve computational efficiency. To this end, a wide variety of paradigms have been explored. Some examples include the use of deep neural networks (NNs) \cite{pan2020deepopf}; graphical and convolutional NNs \cite{Falconer}, \cite{Laurent2}; random forests to resolve voltage magnitudes and phases \cite{Rahman}; and deep reinforcement learning in systems with high renewable penetration \cite{li2021deep}. To further expedite the process, the stacked extreme learning machine (SELM) framework has been explored to bypass the need to learn parameters of a NN \cite{lei2020datadriven}. 
Most akin to our present work are efforts to accelerate the solution of OPF by identifying sets of active constraints. The general philosophy behind this method, and an illustration that the number of ``relevant'' active sets at the optimal solution of an OPF is few, is explored in detail in \cite{misra2019learning}. A more focused explanation and illustration on DC-OPF is given in \cite{DeepMisra}, and the notion that OPF can be reduced to solving a system of linear equations once the active sets are defined is introduced. In particular, a NN-based classifier is developed for determining the most likely ``active sets'' corresponding to the DC-OPF solution. Given a set of manifestations of uncertainty from, e.g., renewables and demand patterns, the classifier returns a set $\mathcal{A}$ of active sets, which should contain the ``optimal active set'' with high probability. This is, by contrast, more efficient than traditional ``ensemble methods,'' in which \emph{all} possible active sets are considered. This general methodology is also followed in \cite{Pascal2021}, on identifying sets of active generators. Additionally, a variant of the active set philosophy was successfully applied by the winners of the ARPA-E GO competition in 2018--2019 \cite{GOcomp}.

With the exception of \cite{Pascal2021}, most of the above and related works are posed from the perspective of a system operator aiming to determine and deploy optimal dispatch. In addition, they are concerned mostly with the primal OPF problem. Furthermore, in those works which do explicitly address ML for pricing of generators, such as \cite{li2021machine}, a map directly to LMPs (or, to be precise, ``LMP spread'') is proposed, rather than solving for LMPs algebraically after intermediate quantities are learned. In our work, we consider the dual problem to DC-OPF, from which locational marginal prices (LMPs) of generators may be computed. This tactic yields an ML scheme that results in LMPs which are faithful to the constraints imposed during OPF, and is therefore of use to energy market participants as well as system operators.

In the remainder of the manuscript, we state two formulations of DC-optimal power flow (DC-OPF), using the latter to construct a system of equations from which all primal and dual variables may be solved. We then outline how LMPs may be computed as a function of these primal and dual variables. The heart of the mansucript is a deep learning scheme that expedites the solution of OPF and computation of LMPs, by classifying a priori which lines in the grid are saturated and incorporating the corresponding thermal constraints, satisfied with equality, into the aforementioned system of equations. The paper concludes with numerical evaluation of this classification, and of the scheme in general. 

\section{Problem formulation}

An optimal power flow (OPF) is the task of finding the economic dispatch, i.e., the least-cost way to generate enough electricity to meet demand.  In the following, we first provide a synopsis of the traditional formulation of a DC-OPF and how it can be rewritten in another equivalent form based on the power transfer distribution factors (PTDFs) \footnote{PTDF generalizes to a more general case of linearized AC-OPF, which we plan to discuss in the extended version of the manuscript.}. In the DC approximation \cite{stevenson1994power}, the ohmic losses in the system are neglected, and the power lines are solely characterized by their susceptances $b_{ij}$. In the present work, we focus on the applications of the DC-OPF to real-time dispatches. It is assumed that unit commitment has been determined a priori, e.g., via security-constrained unit commitment (SCUC) \cite{SCUC}. For the sake of simplicity, we assume that the generators have quadratic costs, with quadratic coefficients $q_g$ and linear coefficients $c_g$. The traditional formulation of the DC-OPF problem is
\begin{align}
\min_{p_g}& \sum_{g} \big(q_g p_g^2 + c_g p_g\big)\label{eq:stdopf1}\,,\\
&{\rm s.t.}\nonumber\\ 
& \forall g: p_g^{\min} \leq p_g \leq p_g^{\max} \,,\\
&\forall i: \sum_{j\sim i}b_{ij}(\theta_i-\theta_j) = \sum_{g \in G_i}p_g - \ell_i\,,\label{eq:pf1}\\
&\forall j\sim i: -f_{ij}^{\max} \le b_{ij}(\theta_i-\theta_{j})\le f_{ij}^{\max}\,,\label{eq:pf2}\\
& \theta_{\rm slack} = 0\label{eq:stdopf2}\,.
\end{align}
where $G_i$ is the set of generators connected to bus $i$ and $j\sim i$ denotes the neighbours of bus $i$. The objective is to minimize the total cost of generation under some operational constraints. Constraint (2) enforces the power flow equations. Constraint (3) ensures that generators do not surpass their capacities when dispatched. Constraint (4) enforces that the power flows must be smaller in absolute value than their thermal limits $f_{ij}^{\max}$. Finally, constraint (5) establishes a reference (or slack bus) to which all voltage phases are compared. 

The PTDF matrix $\bm \Phi$ allows us to directly obtain the power flows in the system from load $\ell_i$ and generator output $p_g$. For the sake of completeness, we briefly show how the PTDF matrix is obtained. Let $\bm B=\bm A\,{\rm diag}(\bm b)\, \bm A^\top$, with $\bm b$ being the vector composed of the line susceptances and $\bm A$ the incidence matrix, defined as
\begin{equation}
A_{ik} =\begin{cases}
-1, & \text{ if line $k$ starts at bus $i$}\,,\\
\phantom{-}1, & \text{ if line $k$ ends at bus $i$}\,,\\
\phantom{-}0, & \text{ otherwise}\,.
\end{cases}
\end{equation}
With these new definitions, Eqs.~\eqref{eq:pf1} and \eqref{eq:pf2} can be rewritten in a matrix form as
\begin{align}
\bm P &= \bm B \bm \theta\,,\label{eq:vec_pf1}\\
\bm F &= {\rm diag}(\bm b)\bm A^\top \bm \theta\label{eq:vec_pf2}\,,
\end{align}
where $P_i=\sum_{g\in G_i}p_g-l_i$ and with the lines being relabeled as $ij\rightarrow k$. $\bm B$ has one zero eigenvalue associated with the eigenvector $1_b$, consisting of a vector of $1$'s for each bus, and hence solving Eq.~\eqref{eq:vec_pf1} for $\bm \theta$ requires the use of a pseudoinverse $\bm B^\dagger$. Finally, injecting the result in Eq.~\ref{eq:vec_pf2}, we obtain
\begin{equation}
\bm F = {\rm diag}(\bm b)\bm A^\top \bm B^\dagger \bm P \equiv \ptdf \bm P\,.
\end{equation}
The pseudoinverse matrix $\bm B^\dagger$ has also one zero eigenvalue associated with the eigenvector $1_b$, and therefore $\sum_i\Phi_{ki}=0$.

Using $\Phi$, it is possible to solve the OPF problem without explicitly computing $\theta_j$. Hence, we reformulate the OPF problem \eqref{eq:stdopf1}-\eqref{eq:stdopf2} as
\begin{align}
\min_{p_g}& \sum_{g} \big( q_g p_g^2 + c_g p_g \big)\,, \label{eq:opf_c}\\
&{\rm s.t.}\nonumber\\ 
&\forall g: \, p_g^{\min} \leq p_g \leq p_g^{\max}\,, \label{eq:opf_pminmax}\\
&\sum_{g} p_g = \sum_{i} \ell_i\,, \\
&\forall k:  -f_k^{\max} \leq \sum_i\ptdf_{ki} \Big(\sum_{g\in G_i}p_g - \ell_i\Big) \leq f_k^{\max}.\label{eq:opf_f}
\end{align}
This reformulation allows to bypass the introduction of dual variables associated with voltage phases, which yields computational benefits to be detailed in the following sections.\\
\\
In what follows, we make the following\\
\textbf{Assumption:} As the unit commitment is assumed to be provided and fixed, we only consider non-committed active (``free'' for simplicity) generators, i.e. $p_g^{\min}<p_g<p_g^{\max}$, and hence no Lagrangian multipliers were assigned to enforce these min/max constraints. Furthermore, the dispatch associated with committed generators is incorporated as ``negative load'' for the buses to which they are associated.

From the DC-OPF problem \eqref{eq:opf_c} - \eqref{eq:opf_f} and the above assumption, we form a partial Lagrangian function.
\begin{align}
\mathcal{L}(p_g,&\lambda,\mu_k,\nu_k)
\doteq
\sum_g\big( q_g p_g^2 + c_g p_g \big) + \lambda \Big(\sum_{g} p_g - \sum_{i} \ell_i\Big) \nonumber\\
&+ \sum_{k} \mu_k \Big[ \sum_{i} \ptdf_{ki} \Big(\sum_{g\in G_i}p_g - \ell_i\Big)- f_k^{\text{max}} \Big]\nonumber \\
&- \sum_{k}  \nu_k \Big[ \sum_{i} \ptdf_{ki}\Big(\sum_{g\in G_i}p_g - \ell_i\Big) +f_k^{\text{max}}\Big]\,.
\label{eqn: DClagrangian}
\end{align}
\\
\\
Once the OPF problem is solved, the locational marginal price (LMP) at every bus is obtained as:
\begin{equation}
\forall i: \lmp_i\equiv \frac{\partial \mathcal{L}}{\partial \ell_i} = -\lambda + \sum_{k} \ptdf_{ki} \big(\nu_k - \mu_k\big)\,,
\label{eq:lmp}
\end{equation}
where the first term represents the energy components of LMPs and the second term represents the congestion component of LMPs. If there is no congestion in the transmission network, then the second term disappears from \eqref{eq:lmp}, as there are no binding power flow limits and their respective dual variables are zero. 

Finally, by construction the solution is \emph{coherent}, i.e. it provides rigorous guarantees for the following key (in)equalities:\\
$\bullet$ \textbf{Revenue Adequacy:} The sum of payments collected by the market from electricity buyers (i.e., loads)  must be equal or exceed the sum of payments made by the market to the electricity sellers (i.e., generators):
\begin{equation}
\sum_{g} \lmp_{i(g)} p_g \le \sum_{i} \lmp_i \ell_i\,,
\label{eq:ra}
\end{equation}
where $i(g)$ is the index of the bus to which the generator $g$ is connected.\\
$\bullet$ \textbf{Cost Recovery:} Each generator does not produce if their marginal cost is higher than their respective LMP; that is, every generator must break even from the participation in the market:
\begin{equation}
\forall g \colon \, q_p p_g^2 + c_gp_g  \leq \lmp_{i(g)}p_g\,,
\label{eq:cr}
\end{equation}
$\bullet$ \textbf{Efficiency via Strong Duality:} Since DC-OPF, as stated, satisfies Slater's conditions, the optimal values of the primal and dual objectives are equal \cite{beck2014introduction}, which indicates the least-cost primal objective value and thus efficiency. As such,
\begin{align}
&\lambda \Big(\sum_{g} p_g -\! \sum_{i} \ell_i\Big) + \sum_{k} \mu_k \Big[\! \sum_{i} \ptdf_{ki} \Big(\sum_{g\in G_i}p_g - \ell_i\Big)- f_k^{\text{max}} \Big]\nonumber \\
&- \sum_{k}  \nu_k \Big[ \sum_{i} \ptdf_{ki}\Big(\sum_{g\in G_i}p_g - \ell_i\Big) +f_k^{\text{max}}\Big]=0\,.
\label{eqn: SDT}
\end{align}

\section{Fast ML approach to solving the OPF problem}

In this section, we present our approach to solving the OPF problem. It provides a fast way of obtaining generation dispatches and LMPs. We call the approach \textit{identification-then-solving} (ITS). We will detail ITS in the following paragraphs, but in short, the procedure amounts to:
\begin{enumerate}
\item{finding a mapping from loads $\ell_i$ to binding line constraints,}
\item{solving a system of linear equations derived from the OPF constraints, and}
\item{recovering the LMPs.}
\end{enumerate}

\subsection{Identification of Binding Line Constraints}

\emph{Binding constraints} are those that are active at the optimal DC-OPF solution. Our approach originates from the empirical finding that there is typically only a small set of lines which are loaded to their maximal capacities $f_k^{\max}$ \cite{DeepMisra}, and that this set does not vary much from one solved OPF instance to another, given the same system parameters and similar load configurations. This is to be expected, as it is well-known that power systems usually have transmission ``bottle-necks'', i.e. portions that tend to be congested (e.g., as the interface between New York City and Long Island and the rest of the NYISO transmission network \cite{NYISO}). Therefore, most of the constraints defined in Eq.~\eqref{eq:opf_f} are never active. From the \textit{complementary slackness} condition  \cite{beck2014introduction}, if a line constraint is not active, then its dual variable, $\mu_k$ or $\nu_k$, is zero.

It should therefore be possible to correctly identify these lines and, moreover, to exploit their properties to gain information about the state of the system. In other words, we seek to find a map from an input to the binding line constraints 
\begin{equation}
\bm \ell\longrightarrow\bigg[\!\!\begin{array}{c}
\mathcal{B}(\bm \mu)\\
\mathcal{B}(\bm \nu)
\end{array}\!\!\bigg]\,,
\end{equation}
where $\mathcal{B}(x) = \{1 : \text{ if } x \neq 0\,,\; 0 : \text{ otherwise} \}$.  In the present work, we limit ourselves to inputs consisting only of the loads $\ell_i$, but one may imagine more advanced inputs. Multiple approaches can be undertaken to find this map. Here, we opt for using a simple deep NN for simplicity of its implementation. 

\subsection{Solving the OPF as a Set of Linear Equations}

We now show that once the identification stage is concluded as described above, the OPF optimization problem is equivalent to solving a system of linear equations (SLE). As previously mentioned in our assumption, from unit commitment, we assume that we know which generators are producing at their full capacity, and we treat them as negative loads. In the following, $p_g$ denotes power outputs of the ``free'' generators, and $\ell_i$ the residual load obtained after the subtraction of committed generators.

From $\mathcal{B}(\bm \mu)$ and $\mathcal{B}(\bm \nu)$, we define $\C_\nu = \{ k\, \colon \, \nu_k \neq 0 \}$ and $\C_\mu = \{ k, \colon \,\mu_k, \neq 0 \}$ which gather the indices of non-zero dual variables associated with binding line constraints to be known. (Note that there is a one-to-one correspondence between these sets and $\mathcal{B}(\bm \mu)$ and $\mathcal{B}(\bm \nu)$.) Assuming these sets to be known, the optimization problem defined by Eqs.~\eqref{eq:opf_c}\,-\,\eqref{eq:opf_f} becomes equivalent to solving the following set of linear equations
\begin{align}
\forall \mu_k^*: \, & \sum_i\ptdf_{ki}\Big(\sum_{g\in G_i}p_g - \ell_i\Big) = f_k^{\text{max}}\,, \label{eq:mu}\\
\forall \nu_k^*: \, & \sum_i\ptdf_{ki}\Big(\sum_{g\in G_i}p_g - \ell_i\Big) = -f_k^{\text{max}}\,, \label{eq:nu}\\
\forall g: \,&\, 2q_g p_g + c_g + \lambda + \!\!\sum_{k\in C_\mu}\! \ptdf_{ki}\mu_k^*- \!\!\sum_{k\in C_\nu}\! \ptdf_{ki}\nu_k^* =0\,,\label{eq:stationarity}\\
& \sum_{g} p_g = \sum_{i} \ell_i.
\label{eq:balance}
\end{align}

Once these equations are solved, LMPs are straightforwardly obtained from $\lambda$, $\mu_k^*$, and $\nu_k^*$ by applying Eq.~\eqref{eq:lmp}. Importantly, \emph{this solution process can be performed significantly faster than solving the optimization problem}. In particular, solving the system of linear equations is already expedient, and benefits from efficient linear algebra utilities in most programming environments. As such, the ``rate-limiting step'' is the classification of binding line constraints, which can be made similarly efficient if a suitable ML methodology is chosen. In our case, to be described below, once the classification NN is trained and performing satisfactorily, the process can immediately be repeated on an ensemble of samples. Furthermore, ITS is clear and intuitive, accessible to practitioners with fundamental knowledge of linear algebra and machine learning.

\section{Numerical Experiments}

In this section, we apply the ITS approach to a standard test case. We detail how we build our data sets and identifier NN. We use our NN identifier to learn the binding line constraints for different unit commitment configurations. This information is then used to obtain the LMPs and generation dispatch outputs. Most notably, we investigate how the results of our ITS method deteriorate when it fails to correctly identify the binding line constraints, highlighting the need for a judicious choice of ML methodology for this crucial step.

\subsection{Generating Data Sets}

We demonstrate our method by applying it to the standard IEEE-118 test case. Unit commitment configurations are obtained from different load configurations corresponding to annual hourly profiles. We extract the 15 configurations that occur the most often. For each of the configurations, a set of 500 samples is obtained by adding noise to each of the individual loads. We use different volatility strengths, with standard deviations of 1\%, 5\%, and 10\% of the nominal loads.  
Out of 15 configurations, we select three of them to present our results. They are chosen to represent various degrees of success: for Config. \#1, our method is accurate in almost every case and performs well for most samples, but starts to fail when volatility is high in Config. \#2; and it fails in several cases even with a relatively low volatility and a large training set  for Config. \#3.
\begin{center}
\begin{tabular}{l c}
&Free generators\\
\hline
Config. \#1&$\{3,5,11,12,18, 30, 34, 40, 42, 43\}$\\
Config. \#2&$\{2,5,12,26,30,39\}$\\
Config. \#3&$\{5,12,14,20,30,37,39\}$\\
\hline
\end{tabular}
\end{center}

\subsection{Implementing the Identification Process}
As we previously mentioned, we decided to use a neural network $\mathbf{NN}_{\bm \psi}$ to identify the binding line constraints
\begin{equation}
\left[\!\!\begin{array}{c}
\bm y^\nu\\
\bm y^\mu\\
\end{array}\!\!\right]
=
\mathbf{NN}_{\bm \psi}(\bm \ell)\,,
\end{equation}
It is trained to minimize the following loss function
\begin{equation}
L_{\psi}=\!\!
\sum_{s}\sum_{k}\!\Big[L_{\rm reg} \big(y_k^{\nu(s)}\!\!\!, \mathcal{B}\big(\nu_k^{(s)}\big) + L_{\rm reg} \big(y_k^{\nu(s)}\!\!\!, \mathcal{B}\big(\nu_k^{(s)}\big)\big)\!\Big],
\end{equation}
where $L_{\rm reg}$ is the cost function for a logistic regression, which is defined as
\begin{equation}
L_{\rm reg}(x,y) = \begin{cases}
-\log(x), & \text{ if } y=1\,,\\
-\log(1-x), & \text{ if } y=0\,.
\end{cases}
\end{equation}

We use a 4-layer NN with 500 units per layer. The last layer uses a sigmoid activation function, which means that the outputs of the neural network lay in $]0,1[$. Once the NN is trained, we use the threshold of $0.5$ to transform the outputs $y_k^{\nu(s)}$ and $y_k^{\mu(s)}$ to binary variables; i.e., if $y_k^{\nu(s)} > 0.5$, set $y_k^{\nu(s)}$ to $1$, otherwise set to $0$, and similar for $y_k^{\mu(s)}$. We use \texttt{Flux.jl} \cite{innes:2018} to implement the NN and train it.

\subsection{Accuracy of the Identification}

Table~\ref{tab:ident} presents the accuracy of our identification process. As explained previously, Config. \#1 was selected because our method performs well for it. Indeed, there are only a few misidentifications, always representing less than 1\% of the binding line constraints, which leads to excellent predictions for LMPs and generation dispatch outputs. Config. \#2 and \#3 have far more misidentifications. Surprisingly, larger training sets tend to increase their number. As we will show in the next subsection, the number of misidentifications is directly related to the mismatch, and binding line identifications based on larger training sets tend to give better market-clearing results. This also means that failing to identify certain binding constraints is more consequential for the ultimate dispatch and LMP predictions than others.
\setlength{\tabcolsep}{4pt}
\begin{table}[h!]
\center
\begin{tabular}{rrrrrrrr}
\hline
&&\multicolumn{2}{c}{Config. \#1}&\multicolumn{2}{c}{Config. \#2}&\multicolumn{2}{c}{Config. \#3}\\

\multicolumn{1}{c}{$\sigma$} & \multicolumn{1}{c}{size} & \multicolumn{1}{c}{training} & \multicolumn{1}{c}{testing}  & \multicolumn{1}{c}{training} & \multicolumn{1}{c}{testing} & \multicolumn{1}{c}{training} & \multicolumn{1}{c}{testing}\\
\hline
1\% & 50 & 0(0.00) & 0(0.00) & 0(0.00) & 34(0.02) & 1(0.01) & 66(0.05)\\
1\% & 100 & 0(0.00) & 0(0.00) & 0(0.00) & 48(0.03) & 0(0.00) & 273(0.20)\\
1\% & 200 & 0(0.00) & 6(0.00) & 0(0.00) & 104(0.06) & 0(0.00) & 342(0.25)\\
5\% & 50 & 0(0.00) & 0(0.00) & 0(0.00) & 16(0.01) & 9(0.03) & 50(0.05)\\
5\% & 100 & 0(0.00) & 0(0.00) & 0(0.00) & 43(0.03) & 0(0.00) & 165(0.14)\\
5\% & 200 & 0(0.00) & 6(0.00) & 0(0.00) & 79(0.05) & 0(0.00) & 232(0.19)\\
10\% & 50 & 0(0.00) & 0(0.00) & 4(0.00) & 19(0.02) & 26(0.05) & 36(0.04)\\
10\% & 100 & 0(0.00) & 0(0.00) & 0(0.00) & 24(0.02) & 6(0.01) & 103(0.01)\\
10\% & 200 & 0(0.00) & 2(0.00) & 0(0.00) & 35(0.03) & 1(0.00) & 133(0.15)\\
\hline
\end{tabular}
\caption{Number of misidentifications, i.e. $y_k^{\nu|\mu(s)}=1$ and $\mathcal{B}\big(\nu|\mu_k^{(s)}\big)=0$ or vice versa, over the training and testing sets. The ratio of misidentifications to number of binding line constraints is given in parenthesis.}\label{tab:ident}
\end{table}

We have thus far discussed results for three representative configurations. 
Table~\ref{tab:ident2} shows the fraction of misidentifications over the testing set for the remaining 12 unit commitment configurations under inspection. On average, their difficulty in terms of training is somewhere between that of Config. \#1 and Config. \#2. Namely, the fraction of misidentifications over the testing set is generally less than 10\%, suggesting that our method is tractable and yields accurate classification results in the vast majority of typical cases. If encountering a limited number of difficult unit commitment configurations, one is advised to increase the size of the training set or perform manual computations after training to ensure the fidelity of the classifications predicted. From a system operations standpoint, the correct course of action will depend on the particular situation, and indeed, it may be worthwhile to consider alternative methods for the computation of generation dispatches and LMPs for such cases. 

\setlength{\tabcolsep}{2pt}
\begin{table}[h!]
\center
\begin{tabular}{rrrrrrrrrrrrr}
\hline
$\sigma$ &\#4&\#5&\#6&\#7&\#8&\#9&\#10&\#11&\#12&\#13&\#14&\#15\\
\hline
1\%& 0.00& 0.00& 0.00& 0.00& 0.00& 0.02& 0.01& 0.00& 0.00& 0.01& 0.01& 0.00\\
5\%&0.04& 0.01& 0.02& 0.00& 0.03& 0.07& 0.02& 0.00& 0.02& 0.06& 0.05& 0.01\\
10\%&0.05& 0.03& 0.04& 0.02& 0.06& 0.08& 0.06& 0.02& 0.06& 0.09& 0.05& 0.05\\
\hline
\end{tabular}
\caption{Fraction of misidentifications over testing set for the 13 unit commitment configurations not presented in Table~\ref{tab:ident}.}\label{tab:ident2}
\end{table}

\subsection{Comparison of the Solutions of the SLE and of the OPF Problem}

\begin{figure*}[t]
\includegraphics[width=\textwidth]{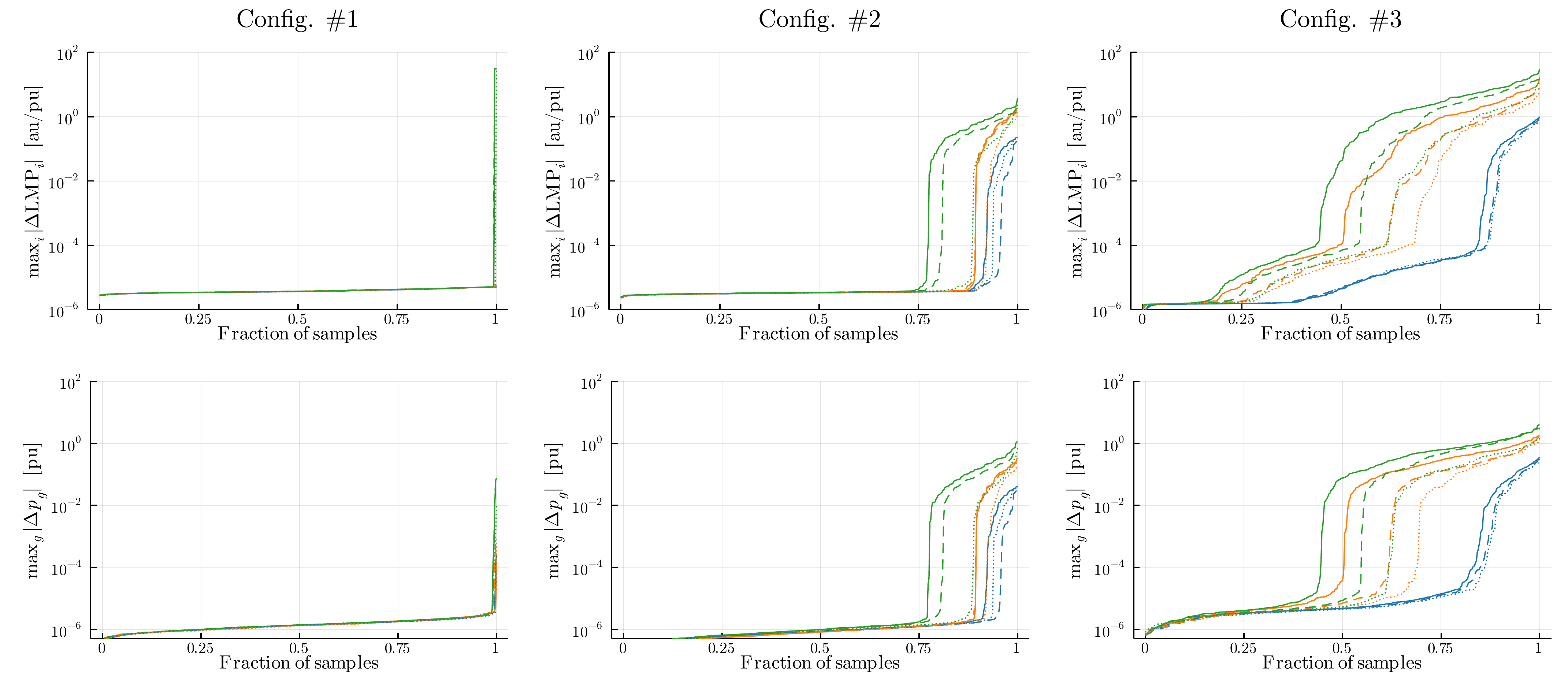}
\caption{Mismatch between obtained LMPs (top) and generator outputs (bottom) for different load volatility standard deviations: 1\% (blue), 5\% (orange) and 10\% (green) of their nominal values, and the training set consists of: 50 (solid), 100 (dashed) are 200 (dotted) samples. Each panel shows the fraction of the testing set that get maximal mismatches smaller than the ordinate.}\label{fig:error}
\end{figure*}

To evaluate the performance of ITS below, we consider the maximum absolute error (MAE) in LMPs and dispatch across each bus. This choice of metric is natural, given that market participants are often interested in the most extreme violations of what is expected in normal operations. Such deviations lead to ``price spikes'' in the LMPs, which trading agents can profit off of if predicted accurately. However, if these spikes are incorrectly anticipated, perhaps as artifacts of the ITS method, it is not as straightforward for trading agents to identify their maximally profitable strategy.

Fig. \ref{fig:error} compares the results of our ITS methods with the ones obtained by solving directly the optimization problem \eqref{eq:opf_c}-\eqref{eq:opf_f}. 
For Config. \#1, the results agree, apart from a very small number of samples when the load volatility is high. For Config. \#2, the higher the load volatility, the higher the mismatches. Training over a large set seems to provide better results. We still obtain excellent results in more than 85\% of cases with a training of 200 samples.
Finally, for Config. \#3, our method is off for a  significant fraction of the testing set.  We hypothesize that misidentifying a binding line with a large associated dual variable affects  the accuracy of our method  more than the total number of misidentifications. If this hypothesis turns out correct, one can take this into account by using $\mu_k^*$ and $\nu_k^*$ as weights in the loss function. This effectively introduces a penalty for each individual misidentification, providing a safeguard against the sensitivity of dispatch and LMP computations to the success of the identification step, or lack thereof.

We analyze if the results we obtain with the proposed method satisfy Eqs.~\eqref{eq:ra} - \eqref{eqn: SDT}. Table~\ref{tab:op_const} shows that \textit{Revenue Adequacy} and \textit{Strong Duality}, as stated in Eq. \eqref{eq:ra} and \eqref{eqn: SDT}, respectively, are almost always satisfied. \textit{Cost Recovery} as stated in Eq. \eqref{eq:cr} is more prone to be violated. When the load volatility is low, \textit{Cost Recovery} is satisfied in most cases. As the volatility increases, the results start to violate \textit{Cost Recovery} in a majority of cases. The magnitude of such violations depend on a chosen unit commitment configuration and is consistent with the current industry practice, that is; such violations can be settled in a out-of-market manner using an uplift mechanism \cite{Kory}. 

\begin{table}[h!]
\center
\begin{tabular}{lccc}
\hline
$\sigma$&Revenue Adequacy & Strong Duality & Cost Recovery\\
\hline
1\% &1.000 & 1.000 & 0.808 \\
5\% &0.999 & 0.997 & 0.352 \\
10\% &0.999 & 0.992 & 0.060 \\
\hline
\end{tabular}
\caption{Fraction of the testing samples (over all 15 unit commitment configurations) that satisfy the key (in)equalities as stated in Eq. \eqref{eq:ra}-\eqref{eqn: SDT}.}\label{tab:op_const}
\end{table}

\section{Conclusion}
In this manuscript, we suggest an identification-then-solving machine learning scheme, consisting of a neural network augmented by traditional linear algebra, to substitute for solution of LMPs. The engine of ITS is to exploit the conditions resulting from balance of load and dispatch, complementary slackness, and stationarity at the point of optimal dispatch. Our scheme relies heavily, almost exclusively, on using a neural network to classify which lines in the transmission grid are binding, and it is observed that the accuracy of both the primal and dual OPF solutions are highly sensitive to the accuracy of the classification step. While the classifier itself performs reasonably well, in order for our proposed scheme to have value, nearly all binding lines must be correctly anticipated, so that all significantly large dual variables are incorporated into the constructed system. Nonetheless, it is encouraging that ITS almost exactly replicates both generator dispatch and LMPs when the classification scheme achieves its goals. In subsequent works, we intend to improve the fidelity of the neural network, namely by taking advantage of the topology of the grid at hand to construct a graphical neural network. Once we achieve the desired level of performance with a GNN, we will also conduct numerical experiments on a larger system, to ascertain the extent to which ITS is robust against various contingencies in power grid operations. Overall, ITS presents a theoretically and mechanically simple alternative to computing dispatch and LMPs through OPF, and such methodology will be valuable to rapid decision-making on the part of both system operators and energy market participants.


\section*{Acknowledgment}

The authors would like to collectively acknowledge with gratitude the ARPA-E PERFORM grant from which this work arose. In addition, R. Ferrando was partially supported by the University of Arizona NSF RTG in Data-Driven Discovery. All authors also thank Daniel Bienstock and Robert Mieth for useful discussions, guidance, and technical assistance at various stages of this work. 

\section*{Appendix: Notations}
Throughout our derivations, matrix $\bm D$ distributes the generators to the buses they are connected to, i.e.
\begin{equation}
D_{ig} = \left\{\begin{array}{l}
1\,, \text{ if generator $g$ is connected to bus $i$}\,,\\
0\,, \text{ otherwise}\,.
\end{array}\right.
\end{equation}

The operative system of equations \eqref{eq:mu}-\eqref{eq:balance} may be more compactly represented in matrix form. Below, $\bm{Q}={\rm diag}(\{q_g\})$ is a matrix of quadratic costs, and $\bm{\Phi}_\nu$, $\bm{\Phi}_\mu$ are reduced PTDF matrices, only accounting for those unsaturated lines in which $\mu_k, \nu_k \neq 0$.
\begin{equation}
\! \! \!
\begin{bmatrix}
\!\!2\bm Q & \! \! \! \! \! \! \! \! \! (\bm{\Phi}_\nu \bm{D})^\top & \! \! \! \! \! (\bm{\Phi}_\mu \bm{D})^\top & \! \! \! \! \bm 1_{\rm g}\\
\bm \Phi_\nu \bm D  &  \bm 0 &\bm 0 & 0\\
-\bm \Phi_\mu \bm D  & \bm 0 &\bm 0 & 0\\
\bm 1_{\rm g}^\top & \bm 0 &\bm 0 & 0\\
\end{bmatrix}
\cdot
\begin{bmatrix}
\bm p\\
\bm \nu^*\\
\bm \mu^*\\
\lambda
\end{bmatrix}
\!=\!
\begin{bmatrix}
-\bm c\\
\bm f_\nu - \bm \Phi_\nu\bm l\\
\bm f_\mu + \bm \Phi_\mu\bm l\\
\sum_i \ell_i\\
\end{bmatrix}
.
\end{equation}
The above matrix may be nearly degenerate if; e.g., two generators share the same ``coordinates'' such as costs. As such, when solving the resulting system of equations, extra care must be taken when solving numerically; for example, one may left-multiply by pseudo-inverse of the coefficient matrix of the variables. Finally, it is worth noting that following our chosen DC-OPF formulation, we do not have to introduce an additional $N_{\rm b} + 1$ equations and unknowns associated with voltage phase angles.



%



\end{document}